\documentclass[12pt,epsf]{article}
\def\be{\begin{equation}}
\def\ee{\end{equation}}
\def\bea{\begin{eqnarray}}
\def\eea{\end{eqnarray}}

\def\be{\begin{equation}}
\def\ee{\end{equation}}
\def\bea{\begin{eqnarray}}
\def\eea{\end{eqnarray}}

\def\case#1/#2{\textstyle\frac{#1}{#2}}
\def\k0{\kappa_{0}}

\begin{document}
\begin{titlepage}

\vspace{.7in}

\begin{center}
\Large
{\bf Power-Law Inflation from the Rolling Tachyon}\\
\vspace{.7in}
\normalsize
\large{ 
Alexander Feinstein
}\\
\normalsize
\vspace{.4in}

{\em Dpto. de F\'{\i}sica Te\'orica, Universidad del Pa\'{\i}s Vasco, \\
Apdo. 644, E-48080, Bilbao, Spain}\\
\vspace{.2in}
\end{center}
\vspace{.3in}
\baselineskip=24pt

\begin{abstract}
\noindent Modeling the potential by an inverse square law in terms of the 
tachyon field ($V(T)=\beta T^{-2}$) we find  exact solution for  spatially flat isotropic universe.
 We show that for $\beta>2\sqrt{3}/3$ the model undergoes power-law inflation.
 A way to construct other exact solutions is specified and exemplified. 
\end{abstract}

\vspace{.3in}

\end{titlepage}

 Several interesting attempts to  reconcile the inflationary paradigm with string theories 
have resulted recently
in much work ( see for example \cite{stringy}). These efforts, however, did not produce a completely 
successful union between
the string theory and cosmology. It is fair to say, at this stage, that string inspired cosmologies
are not a match to standard inflation. Yet, many believe that superstrings
or their generalisations provide an adequate description of fundamental interactions including
gravity. It is therefore important to pursue exploring the connections between the string theories and 
cosmology in order to find as to which, if at all, relic, or aspect of the fundamental theory may
adequately account for the expansion of the universe.  

Much has been written and emphasised about the role of the fundamental dilaton field in the context
of string cosmology. Less is known about tachyon component. This is
 mostly due to the fact that tachyons were considered rather a nuisance in string theory.
Recent developments  in
fluid description of tachyon condensate in bosonic and supersymmetric string theory due
 to Sen \cite{sen1}, \cite{sen2} have 
resulted in enhancing our understanding of the role of the tachyon.
Based on these works, several papers  studying such a fluid in the 
cosmological context \cite{Gibbons}, \cite{Fair}, \cite{mukohyama}, \cite{mazumdar} have appeared most recently.

The main purpose of this article  is to show that under the assumption that the tachyon potential
behaves as an inverse square in terms of the field, the Einstein equations lead to  
{\em simple} exact  analytic solutions
in the  case of spatially flat isotropic universe. These solutions undergo the so-called power-law
inflationary expansion  if the ``slope" of the potential $\beta>2\sqrt{3}/3$, 
and decelerate otherwise. The fact that the rolling tachyon condensate can lead to power law inflation 
may have important consequences in our understanding of cosmology. 

 We also  indicate and exemplify as to how more general solutions can be constructed.
The choice of the inverse square tachyon potential, may at first sound artificial. Nevertheless, we will see
that the potential behaves qualitatively similar to some exact classical potentials derived in the context of
open string field theory. Apart, there is no need in stressing the importance of having a reasonable
exact solution to the coupled tachyon-gravity equations, just to mention that further studies of qualitative
behaviour and
numerical simulations, including the study of density perturbations, can be contrasted against such an exact solution. 

To this end  we consider  spatially flat FRW line element given by:

\begin{equation}
ds^2  = dt^2  - a^2(t)(dx^2  + dy^2  + dz^2 )
\end{equation}

As shown by Sen  \cite{sen1}, \cite{sen2} a rolling tachyon condensate may be described by an effective
fluid with energy density  and pressure given by

\begin{equation}
\rho = {V(T)\over\sqrt{1 - \dot T^2}}
\end{equation}

\be
\label{eos}
p = - V(T) \sqrt{1 - \dot T^2}, 
\ee
here $T$ is  the tachyon field and $V(T)$ the tachyon potential. It is worth mentioning that this
sort of models with an ordinary scalar field were studied in cosmology on phenomenological ground \cite{Arm}. The pressure and the density
of the fluid  may be derived
from a Lagrangian density ${\cal L}=p(\rho)=- V(T) \sqrt{1 - \dot T^2}$. 

The Friedman equation and the entropy conservation equation 
take the form \cite{Gibbons}, \cite{ Fair}:

\be
3H^2 = \rho ={V\over \sqrt{1 - \dot T^2}},
\ee

and
 
\be
\dot\rho = - 3 H (\rho + p)
\ee
We have set $8\pi G=1$, $H$ is the Hubble parameter and  dot means differentiation
with respect to time.

The entropy conservation equation is, in turn,
equivalent to the equation of motion for the tachyon field $T$:

\be
\label{tfe}
{V \ddot T\over 1 - \dot T^2} + 3 H\, V\,\dot T + V' = 0,
\ee
where $V' = dV/dT$.

 Now, substituting 
 $V$ from the Friedman equation together with the expressions for the pressure and the 
density into
the entropy conservation equation  one
can come with the following interesting relation expressing the change of the tachyon in terms
of the Hubble parameter and its derivative:

\be
\label{basic}
\dot T^2=-\frac{2\dot H}{3H^2}
\ee
Below, we will show how this equation can be used to construct variety of exact solutions,
but first  we  derive  solutions with the inverse square potential.

The  equation (\ref{basic}) and the Friedman equation may be  written yet in a different form
\be
\label{basic2}
\dot T=-\frac{2}{3}H'/H(T)^2
\ee
and
\be
\label{frid2}
H'^2 -\frac{9}{4}H^4(T)+\frac{1}{4}V^2(T)=0 
\ee
We now assume that the potential is an inverse square in terms of the tachyon field:

\be
V(T)=\beta T^{-2}, \hspace{.5in} \beta>0
\ee

 Although this potential diverges at $T=0$, it  fairly mimics the  
behavior of a typical potential in the condensate of  bosonic string theory. One expects the
potential to have a maximum at $T\to  0$ and to  die off for a large
field. Recently two groups \cite{gerasimov}, \cite{Kutasov} have given an 
 exact (in string tension) form of 
the classical potential:

\be
V(T) = V_0\left (1+ {T/T_0}\right)\exp{(-T/T_0)},
\ee
Apart from the unphysical divergence at vanishing tachyon in
our toy model, the exact potential (11) has  qualitatively similar behavior.

With this assumption we can see that (\ref{frid2}) has a solution $H \sim T^{-1}$.
Substituting this solution into (\ref{basic2}) we find that the tachyon field is linear in 
time. After some simple algebra we have:
\be
\label{scalefactor}
a=t^n
\ee
along with 
\be
\label{power}
n=\frac{1}{3} +  \frac{1}{6}\sqrt{4+9\,\beta^2}
\ee
The tachyon field has the form:
\be
T=\sqrt{2/3n} \,t
\ee

The condition for inflation for these solutions is that $n>1$. These are
the so-called power-law inflationary solutions \cite{Lucc}, \cite{Hulliwel} and in the 
inflaton driven models are associated with  the exponential potential.
Limitations on $n$ come from the reality
of the equation (\ref{basic}). This imposes the positivity of $n$, and this is why the negative
branch of the equation (\ref{power}) was discarded.
 The same equation prohibits, in general, the so-called super, or pole inflation,  by imposing
 a non-positive sign to the time derivative of the Hubble parameter. From the
equation (\ref{basic}) we immediately find that the inflationary solutions with
$n>1$ correspond to $\dot T^2<2/3$ (cf. \cite{Gibbons}). These models  actually inflate forever.

In terms of the potential parameter $\beta$, the inflation occurs whenever $\beta>2\sqrt{3}/3$.
The range of the parameter $\beta$ for decelerating models is $0<\beta<2\sqrt{3}/3$. In terms
of the power $n$, the last inequality translates into $2/3<n<1$. 

We were lucky of course in choosing the form of the potential in order to solve
the equation (\ref{frid2}) and the rest of the equations exactly.
Given a different potential solving the coupled differential equations
happens to be  rather difficult a task. One may, however, approach this problem
from a different perspective. 
Rather then starting with a given potential, one can
start with a given expansion factor  $a(t)$. The Hubble parameter $H(t)$ is then readily
found  and the tachyon field $T(t)$ can be read from the  equation (\ref{basic}).
The  potential in form of $V(t)$  is then found from the equation:
\be
V=3H^2\sqrt{1-\dot T^2}=3H^2\sqrt{1+\frac{2\dot H}{3H^2}}
\ee 

Inverting  $T(t)$ to get $t(T)$ we finally find $V(T)$.
The main drawback in this scheme is that one may often
finish with  unphysical tachyon potential, the advantage, 
on the other hand, is that  given a known
cosmological expansion, one can figure out the tachyon potential and the field itself.

Alternatively, one
can start with his ``favorable" behaviour for the tachyon. Equation (\ref{basic}) is then
easily integrated to give one the Hubble scale $H^{-1}=\frac{3}{2}\int\dot T^2 dt$.
From here to find the scale factor and the potential $V(T)$ is just a matter of algebra.  

Let us see how it works.
The  case of exponential expansion with the constant Hubble parameter  corresponds to the constant
tachyon and the constant potential. The limiting $t^{2/3}$ behavior corresponds to pressureless dust.
To exemplify  the procedure  for  less straightforward cases we  assume that the scale factor behaves
as
\be 
a(t)=\exp{(mt^n)}
\ee 

After some   algebra we find:

\be
H=mn\,t^{n-1}
\ee

together with

\be
\label{tachyon}
\dot T^2=\frac{2(1-n)}{3mn}t^{-n}  \Longrightarrow T=\gamma t^{(2-n)/2}
\ee

Here $\gamma=\frac{2\sqrt{6(1-n)/mn}}{3(2-n)}$, and the integration constant in (\ref{tachyon}) has
been ignored. We can now invert $T(t)$ to get $t(T)=(T/\gamma)^{2/2-n}$. 
 In terms of $T(t)$, the potential becomes

\be
V(T)=AT^{4(n-1)/(2-n)}\sqrt{B+CT^{-2n/(2-n)}},
\ee
where $A$, $B$ and $C$ are constants expressed in terms of $m$ and $n$.
The special case  $n=2$, should be treated separately, and leads to logarithmically divergent tachyon at $t=0$.
 
The reality condition for the tachyon derivative imposes $0<n<1$ for
$m>0$, and  either $n<0$ or $n>1$ for $m<0$. The sign of the acceleration depends basically
on the  form $mn\,t^n\,(n-1+mn\,t^n)$, depending therefore on $m$ and $n$
one can have  variety of models with different behavior
with respect to inflation. For $n>1$ (negative $m$) the expansion is regular at $t=0$ 
but rather singular
at  $t\to\pm\infty$. These models lead, however, to singular tachyon at $t=0$.
 One can have both non-singular
tachyon and the expansion  at $t=0$ for $1<n<2$. Let us choose, for example, $n=4/3$, then
for negative times the model expands (due to negative $m$, therefore positive $H$) 
from the singularity at  $t\to-\infty$,
starts contracting for $t>0$ and continues to contract towards the singularity at $t\to\infty$.
 As
far as inflation is concerned this model accelerates initially for large negative time, stops
 the acceleration
and decelerates near $t=0$. It finally has an accelerated collapse for large positive time.

To sum up. Assuming a toy model potential for the tachyon field we have shown that this choice leads 
to potentially very interesting power-law inflationary solutions. We have found a simple equation
(\ref{basic}) which allows by starting with a given expansion to find the exact form of the 
tachyon field and its potential.

Some interesting questions remain open. It is well known that in the standard inflaton driven cosmology,
the power-law inflation represents a late time attractor  when the potential 
is exponential \cite{Hulliwel}. It would be interesting to see whether the  power-law
inflationary solutions driven by the inverse square potential of the tachyon have  similar status.
We also are aware that in the inflaton driven expansion, one may re-construct the potential
starting from the behaviour of the density perturbations \cite{jim}, where the so-called
$H(\phi)$ formalism due to Lidsey \cite{jim2} is used. Here one can obviously undertake a similar
task using $H(T)$ as an analog of $H(\phi)$, with the equation (\ref{basic}) being the input for such a study.
Finally, it would be interesting to study the effects of anisotropies and inhomogeneities in this
setting. We hope to be able to address these questions in near future. 

\centerline{\bf Acknowledgments}

This work was supported by the University of the Basque Country Grants UPV 172. 310-GO 2/99 and
The Spanish Science Ministry Grant 1/CI-CYT 00172. 310-0018-12205/2000.

A day after this paper appeared  on arXiv, an overlaping paper of  T. Padmanabhan ``Accelerated
expansion of the universe driven by tachyonic matter" hep-th/0204150 has also appeared.

\end{document}